\documentclass[%
 aps,
 jmp,%
 amsmath,amssymb,
reprint,%
prb,
showpacs
]{revtex4-1}

\usepackage{graphicx}
\usepackage{dcolumn}
\usepackage{bm}
\usepackage{verbatim}

\begin{document}


\title{Geometric blockade in a quantum dot coupled to two-dimensional and three dimensional electron gases} 
%



\author{K. Yamada}
\altaffiliation{Present address: Department of Physics, Kyushu University, 6-10-1, Hakozaki, Higashi-Ku, Fukuoka 812-8581, Japan}
\email{yamada@phys.kyushu-u.ac.jp}

\author{M. Stopa}

\author{T. Hatano}

\author{T. Yamaguchi}

\author{T. Ota}
\affiliation{Tarucha Mesoscopic Correlation Project, ERATO, JST, Atsugi-shi, Kanagawa 243-0198, Japan}

\author{Y. Tokura}
\affiliation{NTT Basic Research Laboratories, Atsugi-shi, Kanagawa 243-0198, Japan}

\author{S. Tarucha}
\affiliation{Department of Applied Physics, University of Tokyo, Hongo, Bunkyo-ku, Tokyo 113-0033, Japan}



\date{\today}

\begin{abstract}
 We fabricated a quantum dot coupled laterally to a two-dimensional electron gas and vertically to a three-dimensional electron gas in order to investigate the eigenstate dependence of tunneling rate to these gases. We observed a bias-dependent ``geometric" current blockade. By tunneling via the asymmetric couplings, population inversion is induced and a dark metastable triplet state is revealed. The metastable state stops the current transport process, suppresses the current and asymmetrically widens the Coulomb diamond. By analyzing the current as a function of source-drain and gate voltage and the magnetic field, we concluded that this effect is due to the geometric shape of the electronic states in the dot and the current is limited by the tunneling rate due to the eigenstates, that is, artificial $\sigma$-coupling and $\pi$-coupling.
 
\end{abstract}


\maketitle 

\begin{figure}
\includegraphics[width=80mm]{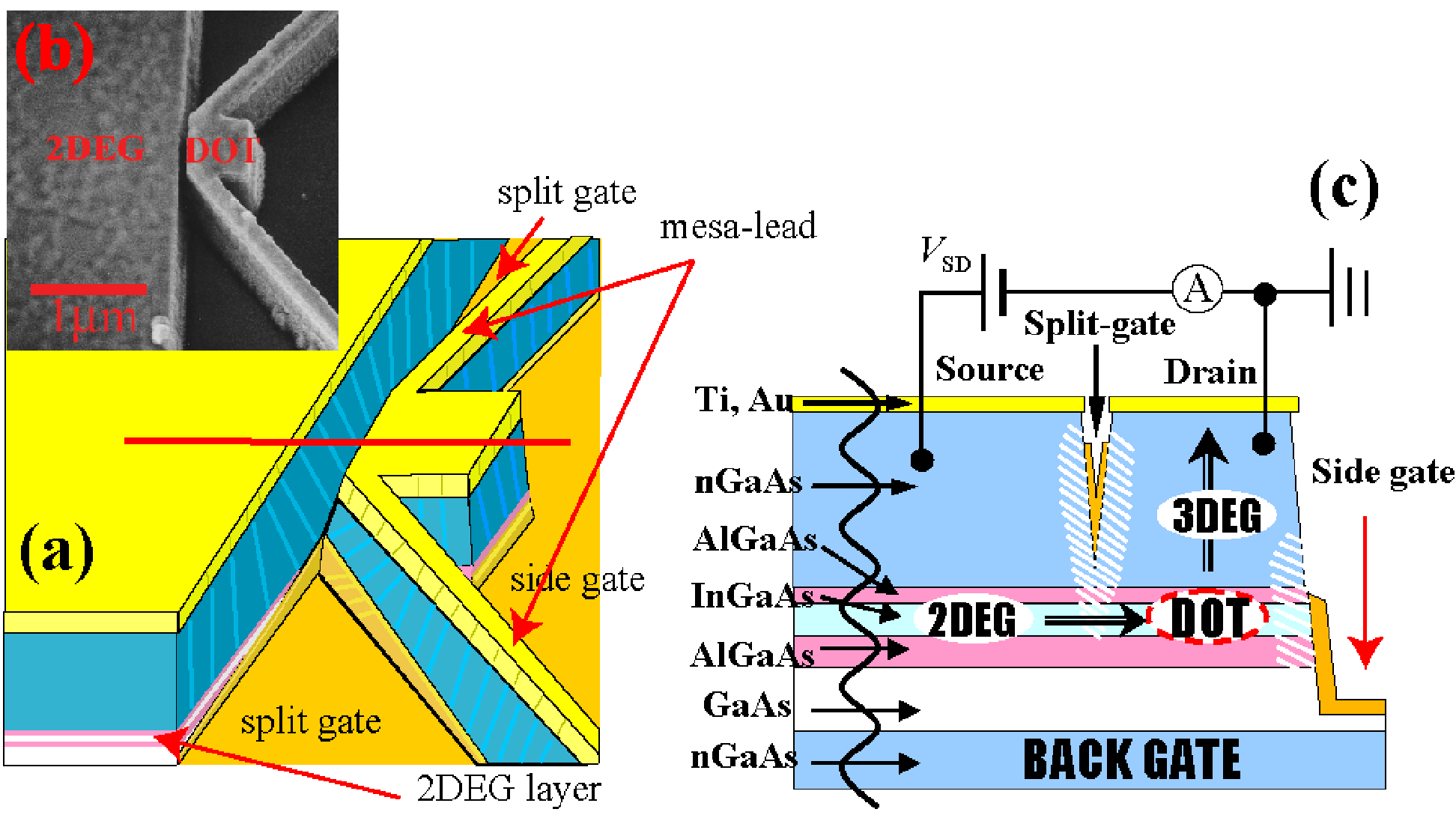}
\caption{\label{fig:Structure}(Color online)(a) Schematic drawing of single artificial atom rectifier. The colors yellow (light gray), orange (light gray), blue (dark gray) and magenta (dark gray) indicate the metal for the electrode, the metal for the gate, n-doped GaAs, and AlGaAs for the tunnel barrier, respectively.(b) SEM image of single artificial atom rectifier. (c) Schematic diagram of cross section where the red (gray) line in (a) indicates the area. The quantum dot is vertically confined in the quantum well fabricated from the AlGaAs(7 nm)/InGaAs(12 nm)/AlGaAs(50 nm) heterostructure. The measured dot is located in the 300 x 300 nm$^{2}$ mesa. Slant lines indicate depletion layers. At the negative bias $V_{SD}<0$, electrons are injected laterally from the 2DEG via the barrier modulated by the split gate to the dot and escape vertically to the 3DEG via the AlGaAs barrier. The current does not flow through the lower thick AlGaAs barrier. The side gate controls the number of electrons in the dot.
 }
\end{figure}

A quantum dot with a few electrons, which is called an ``artificial atom" for the discrete quantum state of electrons confined in a potential, is similar to a real atom \cite{Tarucha96}. When a quantum dot couples vertically to three dimensional electron gases (3DEG), all quantum states are coupled to the 3DEG and atomic-like spectra of all quantum states can be easily obtained \cite{Tarucha96}. On the other hand, when a quantum dot is coupled laterally to two-dimensional electron gases (2DEG), quantum states are not equally coupled to the 2DEG. The spatial probability density of the dot depends on the angular momentum of the electron. The current blockade behaviors due to the angular momentum and the spin have been reported by Ciorga \textit{et al} \cite{Ciorga02}. A combination of such selection rules with a well-defined asymmetry in the couplings between leads and the system can produce complex and potentially useful I-V behavior. Controllable transport in quantum dot systems, which have recently been realized, includes the triple dot Coulomb blockade rectifier \cite{Stopa02,Vidan04} a Pauli blockade in serial double quantum dots \cite{Ono02,Ono04,Liu05}, the coherent manipulation of singlet and triplet states of an $N=2$ double dot \cite{Petta05} and measurements of coupling strengths \cite{Hatano04} and studies of noise \cite{Laird06} in single dot systems with nearby charge sensors. Similar rectification phenomena have been observed in carbon nanotubes \cite{Ishibashi06}, and phase-coherent blockade mechanisms in dots have been proposed, but not yet realized \cite{Michaelis06}. In this paper, we describe a new rectification mechanism that arises, for the electron number $N=2$, in a vertical/lateral GaAs-AlGaAs quantum dot \cite{Yamada03,Yamada05} (see Fig.~\ref{fig:Structure}). This mechanism, which we call a ``geometric blockade," is produced by a single triplet dark state in combination with the designed asymmetry of the dot-lead couplings. 

 The schematic drawing, a scanning electron micrograph (SEM) and a cross section of the device are shown in Figs.~\ref{fig:Structure} (a), (b) and (c), respectively. The dot is separated from a 3DEG lead by a 7 nm AlGaAs barrier and connected to the top ohmic connect, as in standard ``vertical" quantum dots \cite{Tarucha96}. The bottom AlGaAs barrier, however, is very thick and prevents the flow of current to the substrate. Instead, the dot is connected laterally, via a quantum point contact (QPC) tunnel barrier, tuned by two split gates, to a 2DEG between the two AlGaAs barriers. In this study, we fix the voltage of the split gate and control the number of electrons by varying the voltage ($V_G$) of the side gate. This design results in an (approximately) equal coupling of all states to the 3DEG and a coupling to the 2DEG, which is highly state-dependent. The samples are cooled in a dilution refrigerator down to 10 mK, although the electron temperature is estimated to be about 0.1 K.

\begin{figure}
\includegraphics[width=75mm]{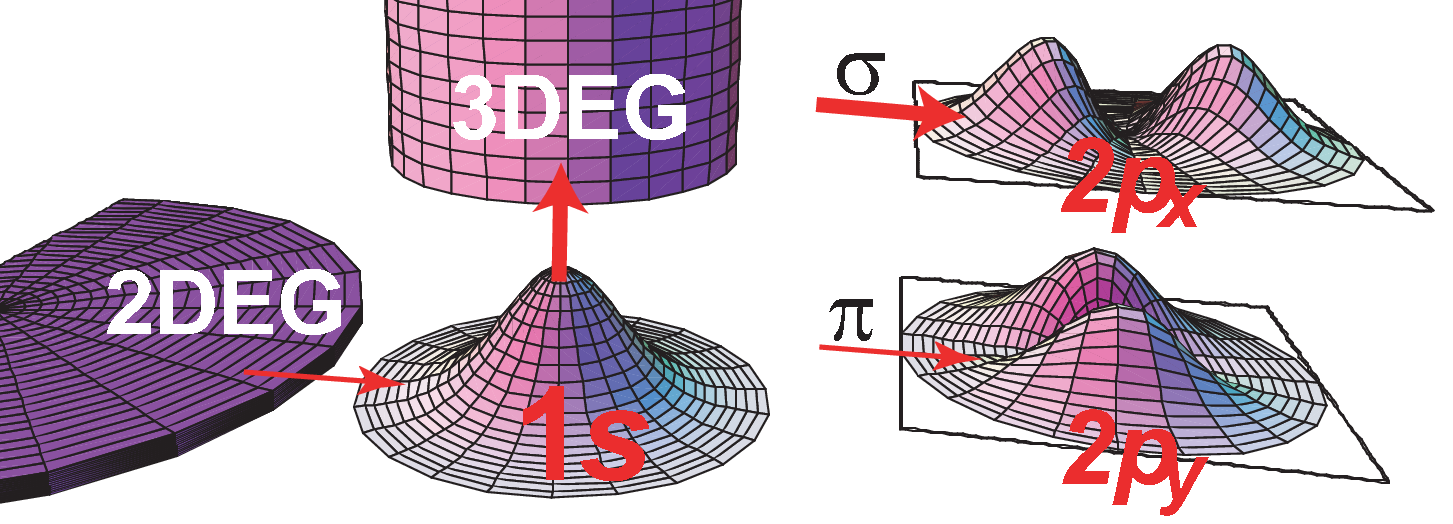}
\caption{\label{fig:image}
(Color online) Schematic of principal single-particle eigenstates in dot. The vertical connections of all states to the 3DEG are approximately the same. However, tunneling to the 2DEG from the $1s$ or $2p_{y}$ state is suppressed by the geometry. The degeneracy of $2p_{x}$ and $2p_{y}$ is broken by the elliptical nature of the potential \cite{Austing99}. The widths of arrows indicate the intensities of couplings. The lateral tunnelings from the 2DEG to the $1s$ and $2p_{y}$ orbitals, are weaker than the tunnelings from 2DEG to a $2p_{x}$ orbital because of a longer effective tunneling distance. The intensity of the vertical tunneling coupling does not depend on the form of the wave function.}
\end{figure}

The main effect of the structural asymmetry is depicted schematically in Fig.~\ref{fig:image}. The potential is roughly parabolic and we therefore use a 2D spectroscopic notation \cite{Tarucha96} uncritically. However, we conclude (see below) that a slight asymmetry reduces the $2p_x$ state relative to the $2p_y$ state, where the x-axis connects the dot to the 2DEG. The dot-2DEG interface is a typical QPC barrier. Therefore, both the potential asymmetry and the energy dependence of tunneling result in state-dependent connectivity to the 2DEG lead. Specifically, $1s$ and $2p_y$ are weakly connected and $2p_x$ is more strongly connected (Fig.~\ref{fig:image}). The couplings of $2p_x$ and $2p_y$ to the 2DEG correspond to $\pi$-coupling and $\sigma$-coupling for real atoms, respectively. The dot-3DEG barrier (not shown in Fig.~\ref{fig:image}) is high and narrow, resulting from the conduction band offset between GaAs and AlGaAs. 

\begin{figure}  
\includegraphics[width=90mm]{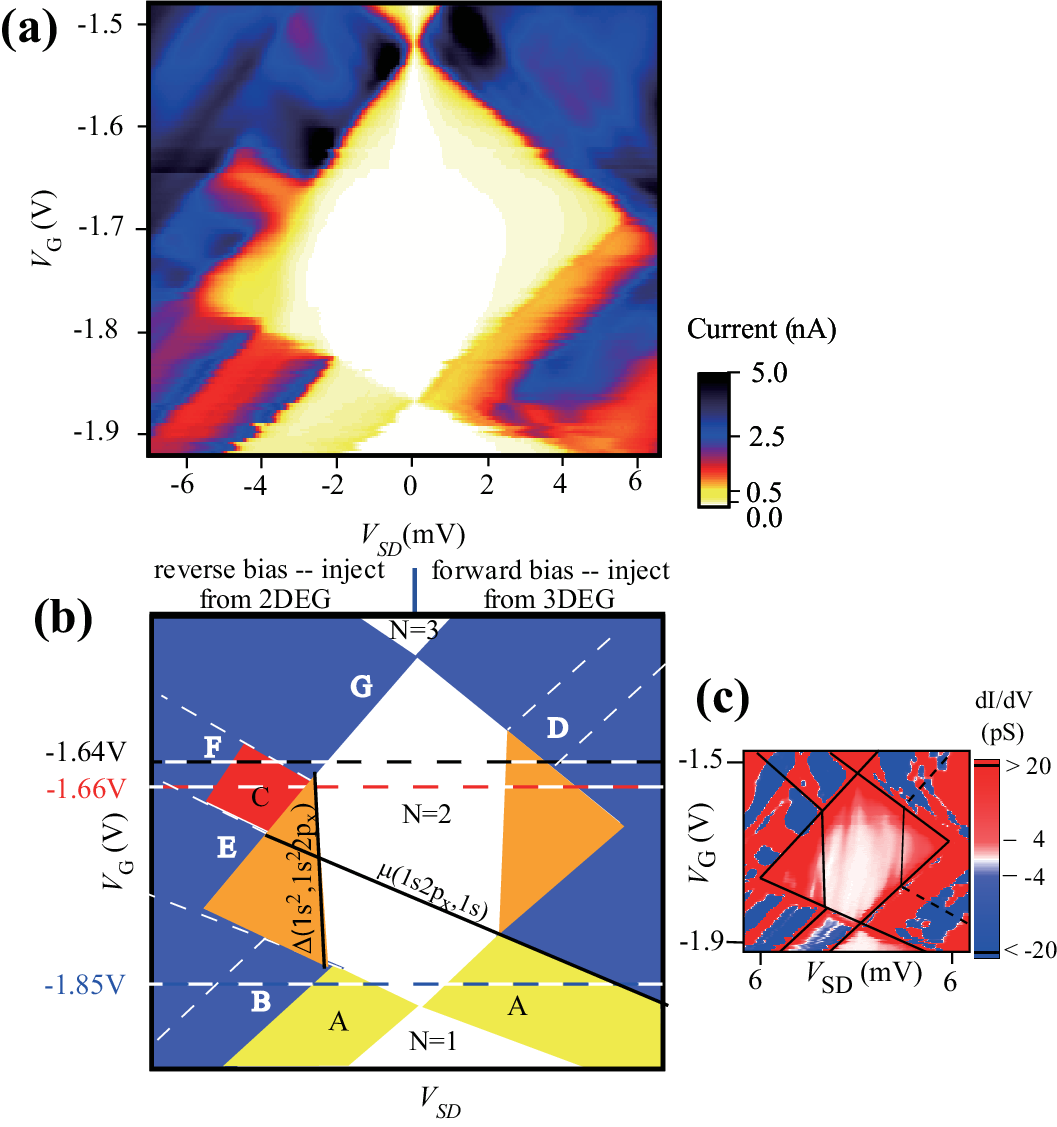}
\caption{\label{fig:diamond} 
(a) $N=2$ Coulomb diamond showing current on color scale versus source-drain and gate voltages. (b) Schematic of (a). For $V_{SD}<0$, injection is from the 2DEG. The blockade at \textbf{C} is due to the dark $1s2p_x$ state, which is stable on the left side of cotunneling line $\Delta (2,2^*)$ and above the ``escape" chemical potential line $\mu (2^*,1)$.  
(c) $dI/dV$ plots of (a). The onset of cotunneling (vertical black lines)is remarkably observed.
}
\end{figure}

\begin{figure}
\includegraphics[width=80mm]{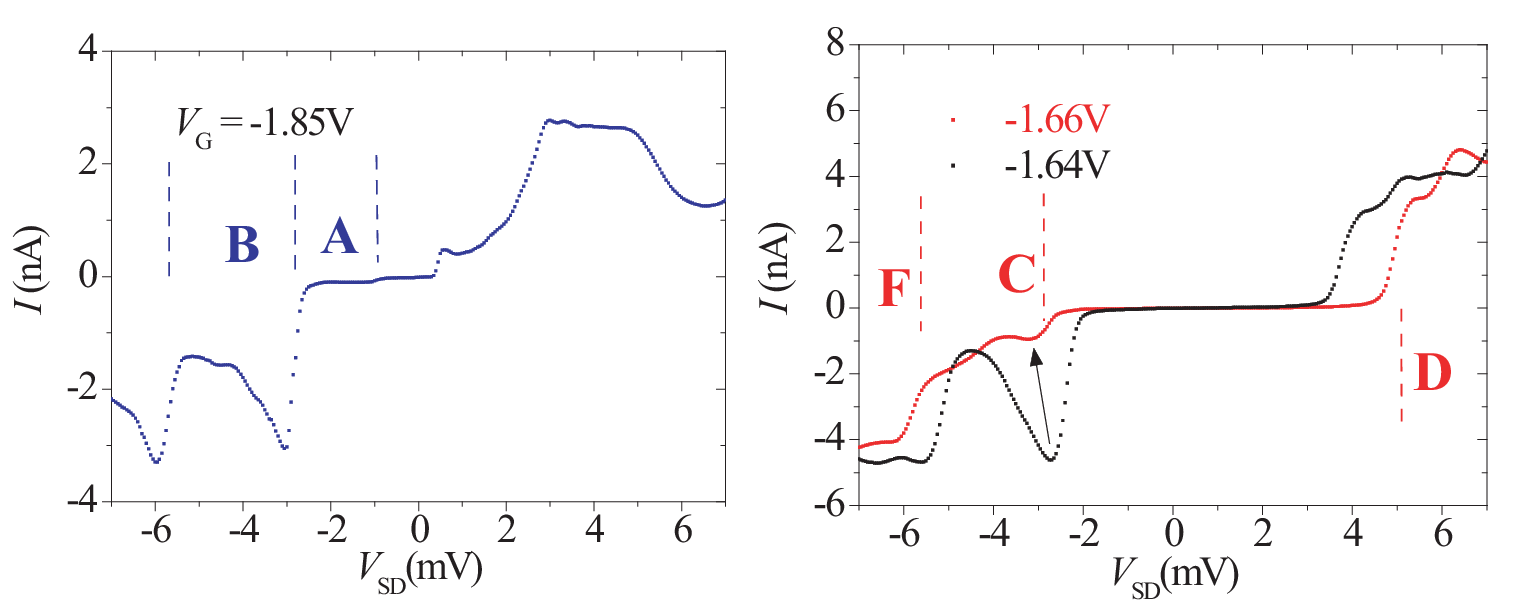}
\caption{\label{fig:IV} 
(Color online) Current vs voltage plots at $V_{G}$= -1.85, -1.64 and -1.66 (V). The markers \textbf{A - F} indicate the same regions in Fig.~\ref{fig:diamond}. The arrow indicates the suppression due to the geometric blockade.
}
\end{figure}

Figure~\ref{fig:diamond} shows the current on a color scale in the $V_{SD}-V_{G}$ plane, showing a Coulomb diamond ($N=2$)  characteristic. The absolute value of currents is obtained in Fig.4 and listed in Table \ref{tab:table1}. Several diamonds (not shown) are observed in the data down to the full depletion of the dot. We show the current vs bias voltage plots for various $V_G$ values in Fig.~\ref{fig:IV}. The difference of tunneling rates between $1s$ and $2p_x$ is indicated in the transport data as follows. For example, in region \textbf{A} the current feature is weak for transition $1s \leftrightarrow 1s^2$ through the $1s$ state. In region \textbf{B}, the current flows via the transition $1s \leftrightarrow 1s2p_x$ through $2p_x$. Then, the current is limited by the tunneling rate of the $2p_x$ state and is higher than that in region \textbf{A}. The $N=2$ diamond displays highly asymmetric transport features including a blockade region (\textbf{C}) located at the upper left diamond edge. In this side (the reverse bias, $V_{SD}<0$ direction), electrons are injected from the 2DEG and exit to the 3DEG. Typically, the upper borders of the $N=2$ diamond demark the threshold for transport between the $N=2$ ground state (GS) and the $N=3$ GS. The aforementioned blockade indicates that the transition $N=2(1s^2)\leftrightarrow N=3(1s^22p_x)$ is suppressed. Note that the blockade region is contact with the cotunneling onset line, where $-eV_{SD}= \Delta (2,2^*)$. Here, $2$ and $2^*$ denote the $N=2$ grand state $1s^2$ and the $N=2$ first excited state $1s2p_x$, respectivly; and $\Delta (2,2^*)=E_{2^*} - E_{2}$ \cite{Franseschi01}, where $E_{i}$ is the energy of state $i$. This cotunneling line is a continuation of the excited state onset line, from the $N=1$ GS to the $N=2$ excited state ($1s\leftrightarrow 1s2p_x$), which abuts the left outer edge of the lower diamond.

\begin{figure}
\includegraphics[width=34mm]{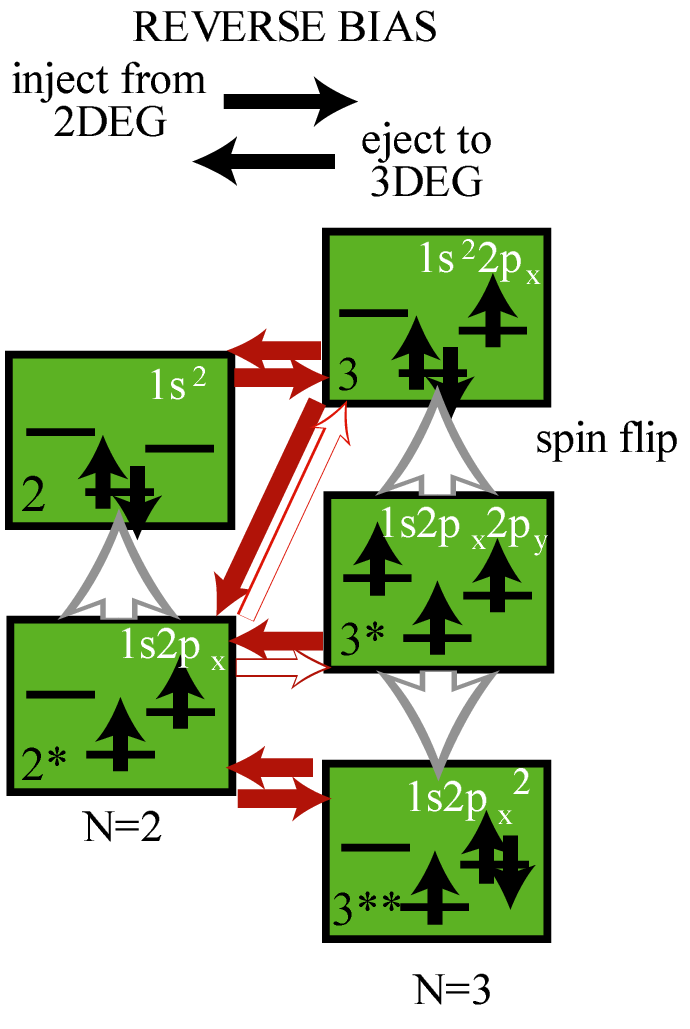}
\caption{\label{fig:Diagram}(Color online) Schematic of states $1s^2$, $1s2p_x$, $1s^22p_x$, $1s2p_x2p_y$ and $1s2p_x^2$ showing strong (filled arrows) and weak (hollow arrows) connections due to single electron tunneling from the 2DEG lead. Note: the diagram is only for the reverse bias ( injection from 2DEG ) case. For example, $1s2p_x\rightarrow 1s2p_x2p_y$ is weak because tunneling from the 2DEG to the $2p_{y}$ level is weak owing to its orientation. However, $1s^22p_x\rightarrow 1s2p_x$ is strong because it is an ejection event to the 3DEG. Thus, when $1s2p_x^2$ is energetically inaccessible at the diamond border, the $1s2p_x$ population strongly increases. Hollow gray arrows denote (weak) relaxation processes, at constant $N$, requiring spin flips. }
\end{figure}

The exchange interaction favors spin alignment and we conclude that the excited $1s2p_x$ state is a spin triplet. We illustrate the configurations of the relevant states and the transitions between them in Fig.~\ref{fig:Diagram} for the case $V_{SD}<0$,with injection from the 2DEG. For $|eV_{SD}| > \Delta (2,2^*)$, $1s2p_x$ can be populated by cotunneling \cite{Franseschi01}. It can also be accessed from the $N=3$ GS in a first-order process \cite{Gustavsson06}. However, $1s2p_x$ cannot decay directly to $1s^2$ without a spin flip process (hyperfine coupling or spin-orbit interaction) or an exchange event with the leads \cite{Hanson05}. Furthermore, the transition from $1s2p_x$ to $1s^22p_x$, in reverse bias, is inhibited by the weak coupling of the $1s$ state to the 2DEG lead. The only other decay path for $1s2p_x$ is via the $N=1$ GS \cite{Ciorga02}, which is energetically accessible when $\mu _{3DEG} < \mu (2^*,1)\equiv E_{2^*}-E_1$ (see Fig.~\ref{fig:diamond}). To summarize, in the region \textbf{C}, the $1s2p_x$ state becomes populated $(1s^2\rightarrow 1s^22p_x \rightarrow 1s2p_x$ and the $1s^2$ state becomes depleted, or $n_{2}/n_{2^*}\rightarrow 0$. The current flows via the $1s2p_x\leftrightarrow 1s2p_x2p_y$ pathway, through the weakly coupled $2p_y$ state. The current is limited by the tunneling rate of $2p_y$. For forward bias, $1s2p_x\rightarrow 1s^22p_x$ is strong because it is an injection event from the 3DEG. Thus in region \textbf{D}, $1s2p_x$ is not metastable. When $1s2p_x$ can decay directly to $N=1$ (its ground state), the population of $1s2p_x$ decreases and that of $1s^2$ increases. Therefore, the outside diamond transport, $N=2\rightarrow 3$, reappears on the left side (region \textbf{E}) of/below the $\mu (2^*,1)$ escape line. In region \textbf{F}($\mu _{2DEG}>\mu (2^*,3^{**})$), the process $1s2p_x\leftrightarrow  1s2p_x^2$ is observed and the transport revivals. The transport process is not limited by the tunneling rate of $\pi $-coupling and the conductance recovers. The main processes that transport the currents in each regions are listed in the upper part of Table I. The chemical potentials required for the transitions are listed in the lower part of Table I. 

\begin{table}  
\caption{\label{tab:table1} \textbf{(upper)} In the regions shown in Fig.~\ref{fig:diamond}(b), the main processes that transport the currents are listed. The tunneling rates $\Gamma$ limiting the main processes are listed at the ``tunnel" column. The spreads of wave functions (Fig.~\ref{fig:image}) indicate the relation $\Gamma _{1s}<\Gamma _{2py}<\Gamma _{2px}$. *The maximum current value which is obtained at the low bias edge near $\mu _{2DEG}=\mu (1,2^*)$. \textbf{(lower)} We treat the Coulomb interaction in the constant interaction model, where the Hartree energy is $C$ and the exchange energy is $K$ \cite{Austing99}. The energies $\varepsilon _{1s}$, $\varepsilon_{2px}$ and $\varepsilon_{2py}$ are the single-electron energies for the $1s$, $2p_x$ and $2p_y$ states, respectively. }
\begin{ruledtabular} 
\begin{tabular} {cccccccc} 
& region & process & tunnel & current\\
\hline
&A&$1s \leftrightarrow 1s^2 $ &$\Gamma _{1s}$ & 0.1nA\\
&B&$1s \leftrightarrow 1s2p_x$ & $\Gamma _{2px}$ & 3nA$^*$\\
&C&$1s2p_x \leftrightarrow 1s2p_x2p_y $ & $\Gamma _{2py} $& 0.9nA \\
&D,E,G&$1s^2 \leftrightarrow 1s^22p_x$ & $\Gamma _{2px}$ & 4nA\\
&F&$1s2p_x \leftrightarrow 1s2p_x^2$ & $\Gamma _{2px}$ & 4nA\\
\end{tabular}
\end{ruledtabular} 
\begin{ruledtabular} 
\begin{tabular} {cccccccc} 
&transition& chemical potential & calculation & \\
\hline
&$1s^2\rightarrow 1s2p_x$ & $\Delta (2,2^*) $ & $\varepsilon _{2px}-\varepsilon_{1s} -K$ & \\
&$1s^2\rightarrow 1s$ & $\mu (2, 1)$ & $ \varepsilon_{1s}+C$ & \\
&$1s2p_x\rightarrow 1s$ & $\mu (2^*, 1)$ & $ \varepsilon_{2px}+C-K$ & \\
&$1s^2\rightarrow 1s^22p_x$&$\mu (2,3)$ & $ \varepsilon_{2px}+2C-K$ & \\
&$1s2p_x\rightarrow 1s2p_x2p_y$& $\mu (2^*,3^*)$&$\varepsilon_{2py}+2C-2K$ & \\
&$1s2p_x\rightarrow 1s2p_x^2$&$\mu (2^*,3^{**})$ & $\varepsilon_{2px}+2C$ & \\
\end{tabular}
\end{ruledtabular}  
\end{table} 


\begin{figure}
\includegraphics[bb=2 260 568 670,clip,width=85mm]{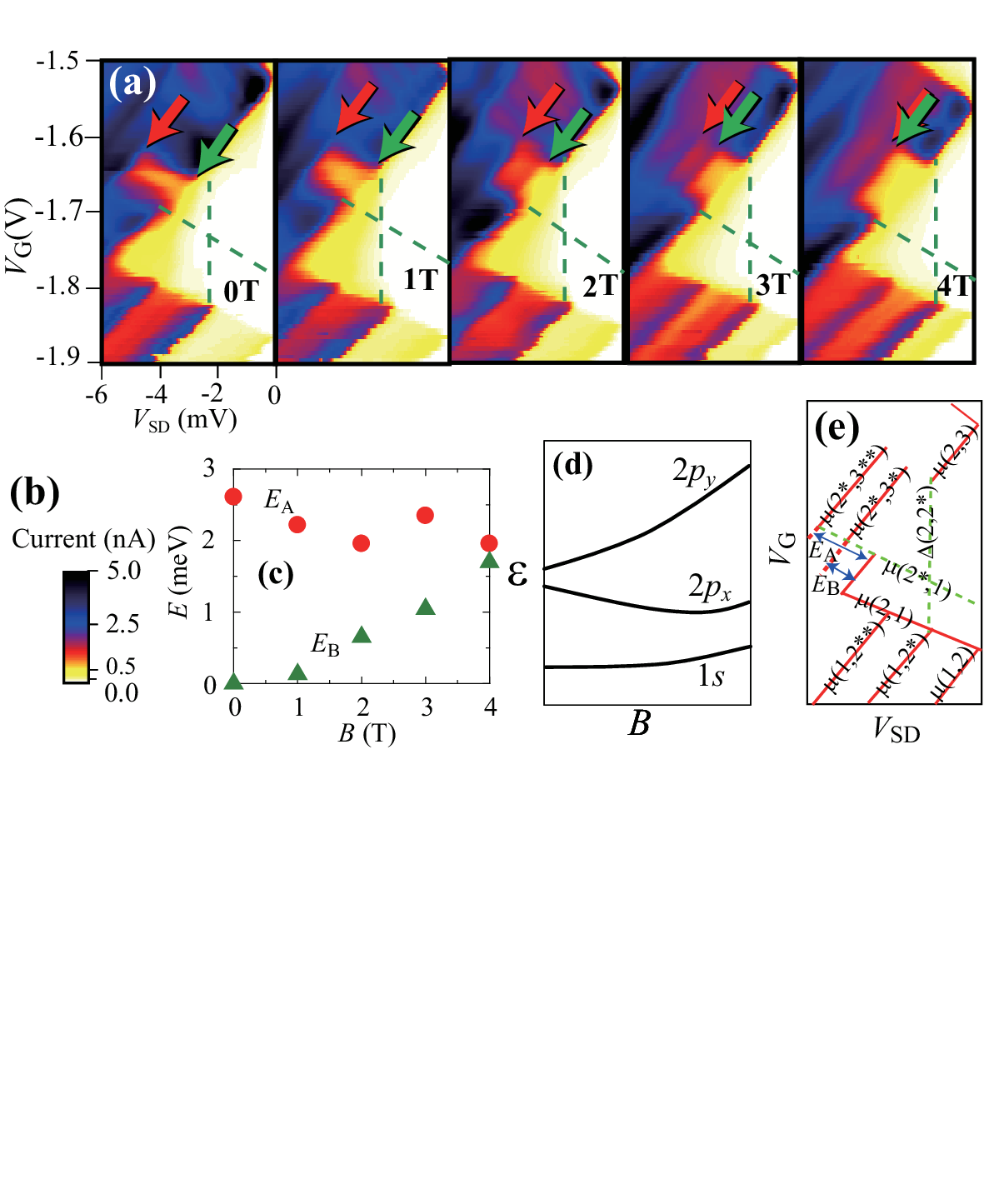}
\caption{\label{fig:mag}
(a) Magnetic field evolution of geometric blockade. The cotunneling line $\Delta (2,2^*)$ and the threshold line for the excited $1s2p_x$ state to decay, via electron escape to the 3DEG, to $N=1$ state, the $\mu(2^*,1)$ are indicated by dashed lines. The excited-state-to-excited state transition features: $1s2p_x \rightarrow 1s2p_x2p_y$(green arrows) and $1s2p_x\rightarrow  1s2p_x^2$ (red arrows) are also shown. The $1s2p_x\rightarrow 1s2p_x2p_y$ transition begins in the blockade (i.e. degenerate with $1s^2 \rightarrow 1s^22p_x)$ at $B=0$ and evolves outwardly from the diamond edge with $B$. This reflects the $B$ dependence of $2p_x$, which fills in the transition (cf. Fig. 5). $1s2p_x \rightarrow 1s2p_x^2$ fills the $2p_{x}$ state, as $1s^2\rightarrow  1s^22p_x$; thus, it does not move with $B$ relative to the diamond edge. (b) Color bar of (a) indicating the current level. Red (gray) lines indicate the maximum of d$I$/d$V$ where the color of (a) changes rapidly. (c)The magnetic field dependence of the distance $E_A$ and $E_B$ which are shown in (e). (d) Schematic of single particle level evolution with magnetic field\cite{Austing99}. States $2p_x$ and $2p_y$ do not degenerate at $B=0$ owing to the ellipticity of the potential. (e) Schematic of (a) and chemical potentials. }
\end{figure}

 An excited-state-to-excited-state transition that degenerates in zero magnetic field is revealed in the magnetic field. Specifically, the energy spacing between $1s2p_x$ and $1s2p_x2p_y$ (see Fig.~\ref{fig:Diagram}) is coincidentally approximately equal to that between $1s^2$  and $1s^22p_x$. Since $2p_y$ is only weakly coupled to the 2DEG, the $1s2p_x\leftrightarrow 1s2p_x2p_y$ feature is weak (i.e. not as dark as the remaining diamond edge where $1s^2\leftrightarrow 1s^22p_x$ is possible), as observed in both the current characteristic and the direct plot of current. The borderline that we identify as $1s2p_x\leftrightarrow 1s2p_x^2$ (red arrow) exists at a higher bias.
The $B$ field evolution of these features (Fig.~\ref{fig:mag}(a)) principally results from the paramagnetic shifting of the $2p_x$ and $2p_y$ single particle levels (Fig.~\ref{fig:mag}(d)); $2p_x$ decreases and $2p_y$ increases. Both the $1s^2\leftrightarrow 1s^22p_x$ and $1s2p_x\leftrightarrow 1s2p_x^2$ transitions empty and fill the $2p_x$ level, whereas for the $1s2p_x\leftrightarrow 1s2p_x2p_y$ transition, the third electron passes through the $2p_y$ level. The GS-to-GS process, $1s^2\leftrightarrow 1s^22p_x$, defines the diamond edge in Fig.~\ref{fig:mag}. Note that the feature resulting from the $1s2p_x\leftrightarrow 1s2p_x^2$ transition does not move much with $B$ relative to the diamond edge, indicating that both the chemical potentials $\mu (2,3)$ and $\mu (2^*,3^{**})$ share the same $2p_x$ magnetic field dependence. However, the $1s2p_x\leftrightarrow 1s2p_x2p_y$ feature, which starts at $B=0$ in the blockade, moves to a higher (more negative) $V_{SD}$ as $B$ increases. Simultaneously, the perfect blockade region (bellow the line $1s^2\rightarrow 1s2p_x2p_y$) widens with increasing $B$, because the single particle energy of $2p_x$, and hence $1s2p_x$, decreases. Therefore, the cotunneling line $\Delta (2,2^*)=V_{SD}$ moves toward the center of the diamond and the escape region bounded by $\mu (2^*,1)=\mu _{3DEG} $ contracts. The distance $E_A$ and $E_B$ are defined as shown in Fig.6(e). The distance $E_A$ does not vary much and $E_B$ increases with $B$. 

We next discuss the application and future of the geometric blockade. The results of this study indicate the feasibility of bulk-state rectifiers using arrays of asymmetrically coupled real atoms. By asymmetrically connecting a single molecule to two electrodes, the geometric blockade due to the wave function will be observed. By vertical/lateral coupling spin-polarized electrons can be injected into a quantum dot and a spin qubit can be initialized \cite{Loss98}.

In conclusion, the properties of the vertical/lateral quantum dot presented here include selective access to atomic states via one (2DEG) lead and non selective access to these states via the other (3DEG) lead. The configuration allows the exploration of discrete-state-mediated transport at the ``atomic" level. One of the most conspicuous features observed in this study is the geometric blockade for the $N=2$ diamond that we have analyzed here. Furthermore, we discussed the applications of the vertical/lateral dot and the geometric blockade.

The authors thank T. Maruyama for helping us in fabricating the device. The authors acknowledge S. Sasaki for the advice in fabricating the device. They also thank W. Izumida, S. Teraoka and T. Sato for various contributions. Part of this work is financially supported by JSPS Grant-in-Aid for Scientific Research S (No. 19104007), MEXT Grant-in-Aid for Scientific Research on Innovative Areas (21102003),and Funding Program for World-Leading Innovative R\&D on Science and Technology(FIRST).


\bibliography{GB_PRB.bib}


%
%

%



\end{document}